\documentclass[aps,prl,twocolumn,amsmath,amssymb,superscriptaddress,longbibliography]{revtex4-2}
\usepackage[breaklinks,colorlinks,bookmarks=false,citecolor=blue,linkcolor=red,urlcolor=blue]{hyperref}
\usepackage{blindtext,enumitem,graphicx,dcolumn,color,xcolor,amsmath,braket,bm,changes,chemformula,cleveref,times}

\graphicspath{{figures/}}
\usepackage{blindtext}
\usepackage{changes}
\usepackage{graphicx}
\usepackage[utf8]{inputenc}
\DeclareUnicodeCharacter{2032}{\ensuremath{\prime}}
\usepackage[english]{fixme}
\usepackage{xcolor}
\DeclareMathAlphabet{\mathbit}{OT1}{cmr}{bx}{it}
\usepackage{wasysym}            % for hexagon
\usepackage{tikz}
\usepackage[caption=false]{subfig}
\crefname{equation}{Eq.}{Eqs.}
\Crefname{equation}{Eq.}{Eqs.}
\crefname{figure}{Fig.}{Figs.}
\Crefname{figure}{Fig.}{Figs.}
\crefname{section}{Sec.}{Secs.}
\Crefname{section}{Sec.}{Secs.}

\begin{document}
\title{Double Supersolid Phase in a Bosonic $t$-$J$-$V$ Model with Rydberg Atoms}

\author{Kuangjie Chen}
		\affiliation{State Key Laboratory of Surface Physics, Institute of Nanoelectronics and Quantum Computing, Department of Physics, Fudan University, Shanghai 200438, China}
            \affiliation{Shanghai Qizhi Institute and Shanghai Artificial Intelligence Laboratory, Xuhui District, Shanghai 200232, China}
        
%\affiliation{\FDU}
\author{Yang Qi}
		\affiliation{State Key Laboratory of Surface Physics, Institute of Nanoelectronics and Quantum Computing, Department of Physics, Fudan University, Shanghai 200438, China}
            \affiliation{Hefei National Laboratory, Hefei 230088, China}
%\affiliation{\FDU}
\author{Zheng Yan}
\email{zhengyan@westlake.edu.cn}
\affiliation{Department of Physics, School of Science and Research Center for Industries of the Future, Westlake University, Hangzhou 310030,  China}
\affiliation{Institute of Natural Sciences, Westlake Institute for Advanced Study, Hangzhou 310024, China}

\author{Xiaopeng Li}
\email{xiaopeng\underline{ }li@fudan.edu.cn}
		\affiliation{State Key Laboratory of Surface Physics, Institute of Nanoelectronics and Quantum Computing, Department of Physics, Fudan University, Shanghai 200438, China}
        \affiliation{Shanghai Qizhi Institute and Shanghai Artificial Intelligence Laboratory, Xuhui District, Shanghai 200232, China}
        \affiliation{Shanghai Research Center for Quantum Sciences, Shanghai 201315, China} 
        \affiliation{Hefei National Laboratory, Hefei 230088, China}

%\affiliation{\SRCQC}

\begin{abstract} 
Recent advances in Rydberg tweezer arrays bring novel opportunities for programmable quantum simulations beyond previous capabilities. 
In this work, we investigate a bosonic $t$-$J$-$V$ model currently realized with Rydberg atoms. Through large-scale quantum Monte Carlo simulations, we uncover an emergent double supersolid (DSS) phase with the coexistence of two 
superfluids and crystalline order. Tunable long-range tunneling and repulsive hole-hole interactions enable a rich phase diagram featuring a double superfluid phase, a DSS phase, and an antiferromagnetic insulator. Intriguingly, within the DSS regime we observe an unconventional thermal enhancement of crystalline order. Our results establish the bosonic $t$-$J$-$V$ model as a promising and experimentally accessible platform for exploring exotic quantum phases in Rydberg atom arrays.
\end{abstract}

\maketitle

{\it Introduction.---} The pursuit of understanding exotic quantum phases has led to significant interest in systems where competing interactions give rise to rich phase diagrams. Rydberg atom arrays, especially those arranged in optical tweezers, offer a versatile platform for simulating quantum many-body systems. The ability to engineer long-range interactions and precise control over individual atoms has facilitated the realization of various quantum models, thanks to the immense experimental progress over the last decade~\cite{bernien2017probing,weimer2010rydberg,semeghini2021probing,ebadi2021quantum,mazurenko2017cold,PhysRevLett.91.090402,gross2017quantum,yan2023emergent}. 

A large theme of Rydberg atom array experiments focuses on spin-$1/2$ models, including transverse field Ising~\cite{Scholl_2021} and quantum XY~\cite{chen2023spectroscopy} models with dipolar or van der Waals interactions. 
These systems have enabled the observation of a variety of exotic phases, such as topological orders~\cite{semeghini2021probing}, quantum scar states~\cite{bluvstein2021controlling}, continuous symmetry breaking~\cite{chen2023continuous} and quantum floating phases~\cite{zhang2025probing}.
Meanwhile, quantum simulations of doped quantum magnetism~\cite{Bohrdt_2021} are of great interest in condensed matter physics, for modeling more complex quantum phenomena. 
A prominent example is the hole-doped spin-$1/2$ model~\cite{PhysRevLett.87.087201,PhysRevB.80.144510,PhysRevA.85.023617,PhysRevLett.132.230401,zhang2024}, whose understanding is expected to be critical to characterizing high-T$_{\rm c}$ superconductivity~\cite{PhysRevLett.132.066002,Jiang_2019,auerbach2012interacting,Jiang_2021}. 

A concrete system to study the interplay of spin fluctuations and hole doping is the bosonic $t$-$J$-$V$ model~\cite{PhysRevLett.132.230401}, recently realized in a Rydberg atom array~\cite{qiao2025realization}.
In this implementation, three Rydberg states of $^{87}\text{Rb}$ atoms---$|61S_{1/2}, m_J=1/2\rangle$, $|60S_{1/2}, m_J=1/2\rangle$, and $|60P_{3/2}, m_J=-1/2\rangle$---encode  spin-up$(\ket{\uparrow})$, spin-down $(\ket{\downarrow})$, and hole $(\ket{h})$, respectively.  The dipolar interactions mediate long-range hole tunneling, while the van der Waals interactions generate both spin-spin and hole-hole interactions. 
The relative interaction strengths are widely tunable by controlling the quantization axis and tweezer array geometry.
Unlike the conventional electronic $t$-$J$ model, where the spin exchange is necessarily antiferromagnetic in the strong coupling limit,  the atomic system permits access to fascinating regimes featuring hole-dynamics in both antiferromagnetic and ferromagnetic spin backgrounds, accompanied by long-range interactions.   
Theoretical investigations of the quantum phases emerging from the complex interplay of spin exchange, hole dynamics and long-range interactions are crucial for advancing quantum simulations in this experimental platform.
 
In this Letter, we present a first study of the ground-state phase diagram of the bosonic $t$-$J$-$V$ model on the square lattice using quantum Monte Carlo. Under strong hole-hole repulsion, the system exhibits conventional antiferromagnetic (AFM) and double superfluid (DSF) phases at small and large tunneling ($t$) limits, respectively. 
At intermediate $t$, we uncover an intriguing double supersolid (DSS) phase—an exotic state characterized by spontaneous breaking of the lattice symmetry and two $U(1)$ symmetries. This DSS phase emerges from the intricate interplay between long-range tunneling, hole-hole repulsion and strong spin-hole couplings. 
Crucially, the long-range interactions inherent to the Rydberg platform stabilize the DSS phase by frustrating domain-wall formation, in contrast to nearest-neighbor Hubbard-type models on the square lattice where supersolids are unstable to phase separation~\cite{PhysRevLett.94.207202}. 
Further experimental studies of DSS phenomena are expected to contribute strongly to the fundamental understanding of doped quantum magnetism, of direct relevance to high-T$_{\rm c}$  superconductivity~\cite{RevModPhys.78.17,mazurenko2017cold,koepsell2019imaging,Brown_2019,koepsell2021microscopic,Jiang_2021,Bohrdt_2022,xu2023frustration,Lebrat_2024,xu2025neutral}. 

\begin{figure}[htb]
\includegraphics[width=0.99\linewidth]{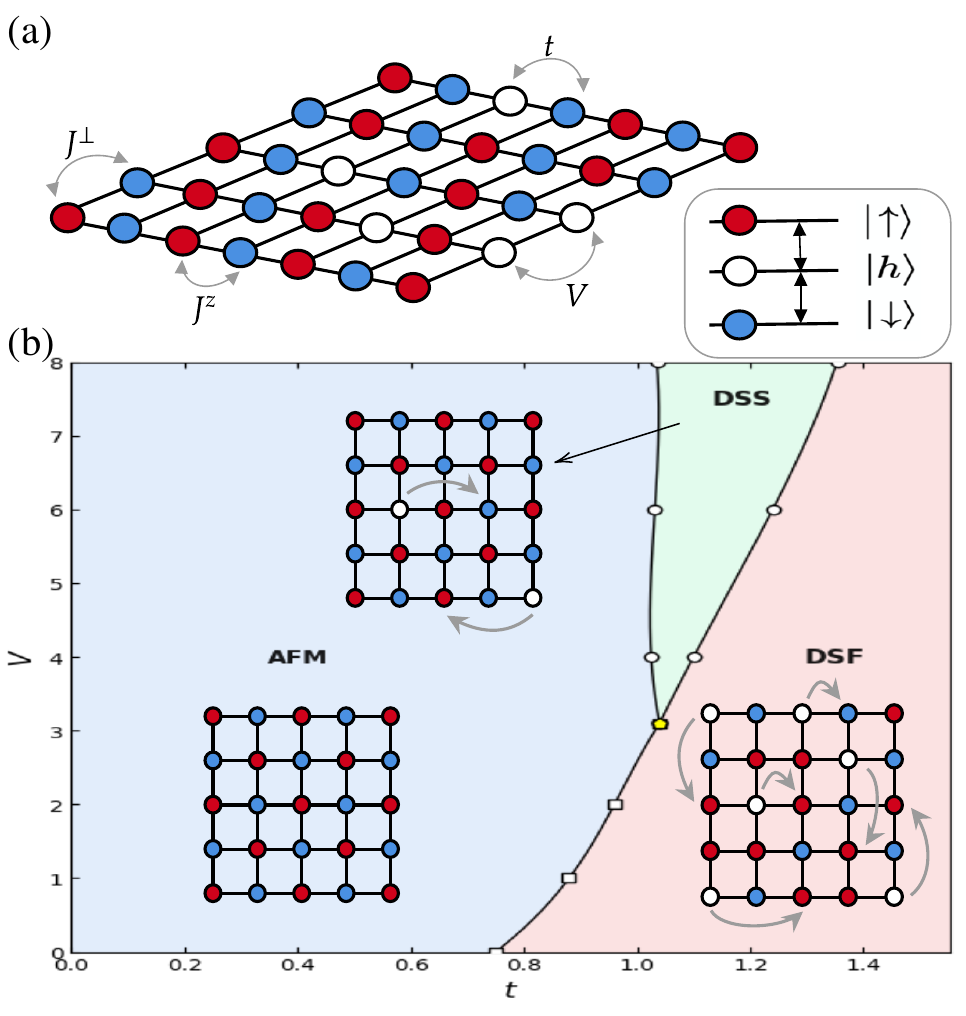}
\caption{(a) Illustration of the bosonic $t$-$J$-$V$ model on the square lattice. The spin-1/2 and hole degrees of freedom are encoded in three distinct Rydberg states, represented by red ($\ket{\uparrow}$), blue ($\ket{\downarrow}$), and white ($\ket{h}$) circles. Effective tunneling $t$, spin-spin interactions $J^{z}$, $J^{\perp}$, and hole-hole interactions $V$ emerge from the interactions among these Rydberg states. (b) Ground-state phase diagram of the $t$-$J$-$V$ model in the $t$-$V$ plane at fixed parameters $J^{z}=4$, $J^{\perp}=-1$, and $\mu=0$. The phase diagram contains an AFM phase (blue region), a double superfluid (DSF) phase (red region), and a double supersolid (DSS) phase (green region). Insets provide schematic illustrations of each phase, with solid lines indicating phase boundaries.}
\label{fig:fig1}
\end{figure}

{\it Model.---} We investigate the bosonic $t$-$J$-$V$ model describing Rydberg atoms on the two-dimensional (2D) square lattice with the following Hamiltonian~\cite{qiao2025realization},
\begin{equation}
\begin{aligned}&\hat{H}_{tJV}=\hat{H}_{t}+\hat{H}_{J}+\hat{H}_{V},\\&\hat{H}_t=-\sum_{i<j}\sum_{\sigma=\downarrow,\uparrow}\frac{t_\sigma}{r_{ij}^3}\left(\hat{a}_{i,\sigma}^\dagger\hat{a}_{j,h}^\dagger\hat{a}_{i,h}\hat{a}_{j,\sigma}+\mathrm{~h.c.~}\right),\\&\hat{H}_{J}=\sum_{i<j}\frac{1}{r_{ij}^{6}}\left[J^{z}\hat{S}_{i}^{z}\hat{S}_{j}^{z}+\frac{J^{\perp}}{2}\left(\hat{S}_{i}^{+}\hat{S}_{j}^{-}+\mathrm{h.c.~}\right)\right],\\&\hat{H}_{V}=\sum_{i<j}\frac{V}{r_{ij}^{6}}\hat{n}_{i}^{h}\hat{n}_{j}^{h} -\mu \sum_{j} \hat{n}_{j}^{h} .\end{aligned}
    \label{eq:hamiltonian}
  \end{equation}  
The effective spin-1/2 and hole states (Fig.~\ref{fig:fig1}(a)) correspond to three Rydberg atomic levels with $r_{ij}$ denoting the distance between site $i$ and $j$. 
The coupling $t_{\sigma}$ describes dipolar tunneling between the spin states ($\ket{\uparrow}, \ket{\downarrow}$) and the hole state $\ket{h}$, with the two corresponding to Rydberg states of opposite parity. The $J^{z}$ and $V$ terms represent the diagonal van der Waals interactions between Rydberg pairs, while the $J^{\perp}$ term corresponds to the off-diagonal van der Waals interactions between $|nS, (n+1)S\rangle$ pairs.
The Schwinger boson operators, $\hat{a}_{j,h}^\dagger$ and $\hat{a}_{j,\sigma}^\dagger$, create a hole and a spin $\sigma$ on site $j$, respectively.  The local hole number and spin number operators at site $j$ are given by $\hat{n}_{j}^{h}=\hat{a}_{j,h}^\dagger \hat{a}_{j,h}$ and $\hat{n}_{j}^{\sigma}=\hat{a}_{j,\sigma}^\dagger \hat{a}_{j,\sigma}$.
The spin-$1/2$ operators are defined as
$\hat{S}_{j}^{\gamma} = \tfrac{1}{2}\sum_{\alpha,\beta} 
\hat{a}_{j,\alpha}^{\dagger}\,\sigma^{\gamma}_{\alpha\beta}\,\hat{a}_{j,\beta},$
where $\sigma^{\gamma} \in \{\sigma^{x}, \sigma^{y}, \sigma^{z}\}$ are the Pauli matrices and $\alpha,\beta$ denote the spin indices $\uparrow,\downarrow$.
 The $\mu$ term is the chemical potential and we focus at $\mu=0$ in this study. Due to the local hard-core constraint of the Schwinger bosons, $\hat{n}_{j}^{\uparrow} + \hat{n}_{j}^{\downarrow} + \hat{n}_{j}^{h} =1$, the Hamiltonian~\eqref{eq:hamiltonian} respects two global $U(1)$ symmetries, corresponding to the conservation of hole dopants, $\hat{N}_{h}=\sum_{j} \hat{n}_{j}^{h}$, and total magnetization, $\hat{S}^{z}_{tot}=\sum_{j} \hat{S}_{j}^z$. 
 The non-equilibrium quantum dynamics of the Hamiltonian has been explored in recent Rydberg experiments~\cite{qiao2025realization}, focusing on the parameter regime of $|J^{z}/J^{\perp}|<1$, $V<0$. 
 In this work, we mainly investigate the thermal equilibrium or ground state physics in the regime of $|J^{z}/J^{\perp}|>1$, $V>0$, which we expect to guide further experimental progress.

 In our simulation, we set $t_{\uparrow}=t_{\downarrow}=t$, $J^{z}=4$ and $J^{\perp}=-1$ (energy unit), for which the model is sign-problem-free~\cite{PhysRevLett.132.230401}. We focus on repulsive hole-hole interactions ($V\geq0$) and explore the phase diagram using the stochastic series expansion (SSE) QMC method with the directed-loop algorithm~\cite{PhysRevE.68.056701, PhysRevE.67.046701,merali2024stochastic}. 
Our simulations are performed on $L \times L$ square lattices with periodic boundary conditions (PBC), while setting the inverse temperature $\beta = L$ to probe ground state properties~\footnote{This choice ensures that the simulation temperature $T=1/\beta$ remains below the finite-size gap $\Delta \sim L^{-z}$, where $z$ is the dynamical critical exponent. In the $t$–$J$–$V$ model, long-range interactions typically lead to larger finite-size gap ($z \leq 1$), so setting $\beta = L$ is more than sufficient to probe the ground state.}. The interactions are truncated at $r_c = 2$ (in units of lattice spacing) for the $r^{-6}$ terms and at $r_c = 4$ for the $r^{-3}$ terms~\footnote{We find that a dipolar truncation of $r_c = 4$ is sufficient to capture the relevant long-range physics of the resulting phases, as verified by additional simulations performed at $r_c = 6$ in the Supplemental Material.}.
The simulations present significant challenges due to the parametric difference between the van der Waals and dipolar interactions~\cite{PhysRevA.111.L011305}, as well as the suppression of quantum fluctuations in the small-$t$ regime.
We employ annealing techniques~\cite{aranson2001dynamics,mitra2018quantumquenchdynamics,yan2023emergent,yan2023quantum} 
and develop QMC algorithms for the bosonic $t$-$J$-$V$ model that are efficient across a wide parameter range. 
The consequent phase diagram is shown in Fig.~\ref{fig:fig1}(b). 
Our results reveal: (i) an antiferromagnetic phase and a double superfluid phase 
(spin in-plane order $+$ hole-superfluidity) 
separated by a first-order transition at $V \lesssim 3.0$, 
(ii) an emergent double supersolid phase 
 (coexisting ferromagnetic in-plane and antiferromagnetic out-of-plane spin orders $+$ hole superfluidity)
between these phases when $V \gtrsim 3.0$, 
 with its stable region widening systematically as $V$ increases.
At finite temperature (simulated with $\beta \ll L$), the DSS phase displays a striking non-monotonic response with increasing temperature. We attribute this to a “thermal compensation ordering” mechanism.

{\it AFM-to-DSF transition.---}
To characterize different phases, we consider the AFM order parameter $|m_z| = \frac{1}{N} \langle | \sum_{j} e^{i\boldsymbol{Q}\cdot\mathbf{r}_j} \hat{S}^{z}_{j} | \rangle$ and the spin structure factor $\textstyle S(\boldsymbol{Q})=\frac{1}{N}\sum_{j,k}e^{i\boldsymbol{Q}\cdot(\mathbf{r}_j-\mathbf{r}_k)}\langle \hat{S}^z_j \hat{S}^z_k\rangle$ at the ordering vector $\boldsymbol{Q}=(\pi,\pi)$. We note that for the hole $\bra{h_{j}}\hat{S}^{z}_{j} \ket{h_{j}} = 0$. The AFM phase that breaks the lattice translation symmetry is indicated by a non-vanishing $|m_z|$ (or $S(\pi,\pi)/N$) in the thermodynamic limit. 
The superfluidity (associated with the $U(1)$ symmetry breaking of holes) is distinguished by a nonzero superfluid density $\rho_{s} = L^{2-d}\langle W_x^{2} + W_y^{2}\rangle/ 2d \beta t$, $d$ is the dimension of the system and $W_{x/y}$ are the winding numbers of bosonic holes' worldlines in the $x$- and $y$-directions~\cite{PhysRevB.36.8343}.

\begin{figure}[htb]
\centering
\includegraphics[width=0.99\linewidth]{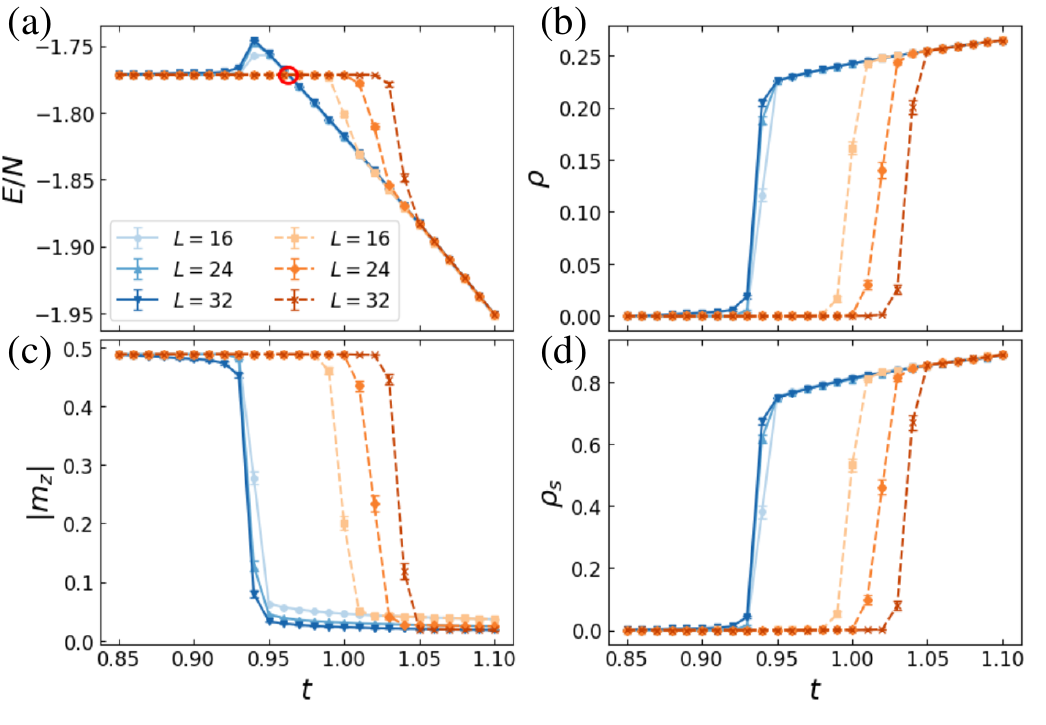}
\caption{(a) Ground-state energy per site $E/N$, (b) hole density $\rho$, (c) AFM order parameter $|m_z|$, and (d) superfluid density $\rho_s$ as functions of $t$ for various system sizes ($L=\beta$). The legend is shown only in (a) since all panels use the same symbols. Blue solid (yellow dashed) lines denote forward (backward) annealing.  The red circle in panel (a) marks the discontinuity point in the derivative of the minimum energy identified from the hysteresis. ($V=2$).}
\label{fig:fig2}
\end{figure}

During our numerical simulations, we encountered challenges with conventional QMC methods when dealing with strong first-order transitions between the AFM and DSF phases. The performance of the directed-loop algorithm is further compromised at small $t$, where weak quantum fluctuations suppress off-diagonal operators in imaginary time, restricting the directed-loop updates to short local moves and thus reducing sampling efficiency. To overcome these challenges, we employed the annealing method~\cite{yan2023emergent}, starting with a large-$t$ value in the DSF region and gradually decreasing $t$ during the simulation. Once in the AFM region, we reversed the process by incrementally increasing $t$. The results, shown in Fig.~\ref{fig:fig2}, reveal that the forward and backward annealing processes yield different outcomes in the intermediate region. This hysteresis loop indicates that the simulation trajectories retain memory of the initial state. Combined with the discontinuous jumps in hole density, $\rho = \frac{1}{N}\langle\sum_j \hat{n}_j^h\rangle$, as $t$ varies, these results strongly suggest a first-order phase transition between the AFM and DSF phases. 
We thus identified the transition point by locating the kink where the derivative of the minimum energy shows a discontinuity (Fig.~\ref{fig:fig2}(a)).

{\it DSS phase.---}
We now focus on the large-$V$ region. Our previous results show that increasing the tunneling $t$ causes holes to proliferate coherently, eventually destroying the AFM order in the small-$V$ region. To stabilize both AFM order and superfluidity simultaneously, we increase $V$, creating strong hole-hole repulsion that suppresses the hole density. This leads to the emergence of a DSS phase characterized by the crystalline order and breaking of two $U(1)$ symmetries.

We set the parameter $V=8$, which gives a strong blockade effect to the hole. The QMC simulations are still performed via the annealing protocol. The hysteresis behavior that happened at weak repulsion regime now disappears. 
In addition to the spin structure factor and superfluid density, 
we measure the fidelity susceptibility, which provides a direct method for detecting quantum phases and phase transitions. For a general Hamiltonian $\hat{H}(\lambda)=\hat{H}_{0} + \lambda \hat{H}_{1}$, where $\lambda$ is the parameter driving the phase transition, the fidelity susceptibility at finite temperature is defined as~\cite{PhysRevE.76.022101,PhysRevLett.103.170501,PhysRevB.81.064418,PhysRevX.5.031007}:
\begin{equation}
\chi_F(\lambda)=\int_0^{\beta/2}\tau[\langle\hat{H}_1(\tau)\hat{H}_1\rangle-\langle\hat{H}_1\rangle^2]d\tau,
\end{equation}
where $\langle \cdot\cdot\cdot\rangle$ denotes the thermal average at inverse temperature $\beta$ and $\hat H_1(\tau)=e^{\tau\hat H}\,\hat H_1\,e^{-\tau\hat H}$. 
In our SSE simulations, the tunneling strength $t$ serves as the driving parameter, and $\chi_F$ is extracted by monitoring the $\hat{H}_t$ operators during the imaginary time evolution.

\begin{figure}[htb]
\includegraphics[width=0.92\linewidth]{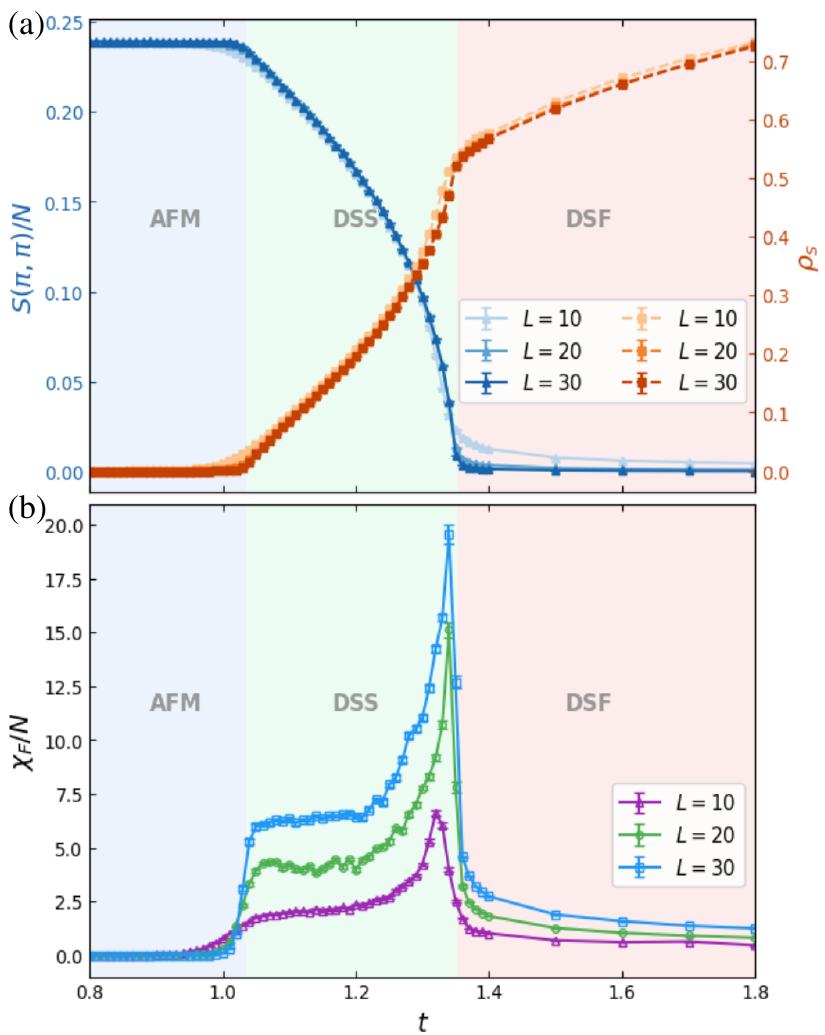}
\caption{(a) Structure factor $S(\pi,\pi)/N$ (blue solid lines) and superfluid density $\rho_s$ (yellow dashed lines) as functions of $t$ for various system sizes ($L=\beta$). (b) Fidelity susceptibility per site $\chi_{F}/N$ , illustrating the AFM-to-DSS and DSS-to-DSF transitions. ($V=8$).}\label{fig:fig3} 
\end{figure}

The results are shown in Fig.~\ref{fig:fig3}.
 The first-order transition that separates the AFM and DSF phases at weak hole-repulsion, is now replaced by an intermediate DSS phase in the middle from $t \simeq 1.035$ to $t \simeq 1.354$, where both of $S(\pi,\pi)/N$ and $\rho_s$ are finite.
The fidelity susceptibility per site $\chi_{F}/N$ provides additional support for the phase transition. As shown in Fig.~\ref{fig:fig3}(b), it approaches zero deep inside both the AFM and DSF phases, but exhibits a sharp increase at the AFM-to-DSS transition, forming a clear plateau throughout the DSS phase. This plateau indicates the increased sensitivity of the DSS to the variations of the tunneling strength $t$. As $t$ increases further, $\chi_{F}/N$ shows a divergent peak at the DSS-to-DSF transition with increasing system size. This behavior is fully consistent with our order parameter results.

\begin{figure}[htb]
    \centering
    \includegraphics[width=0.99\linewidth]{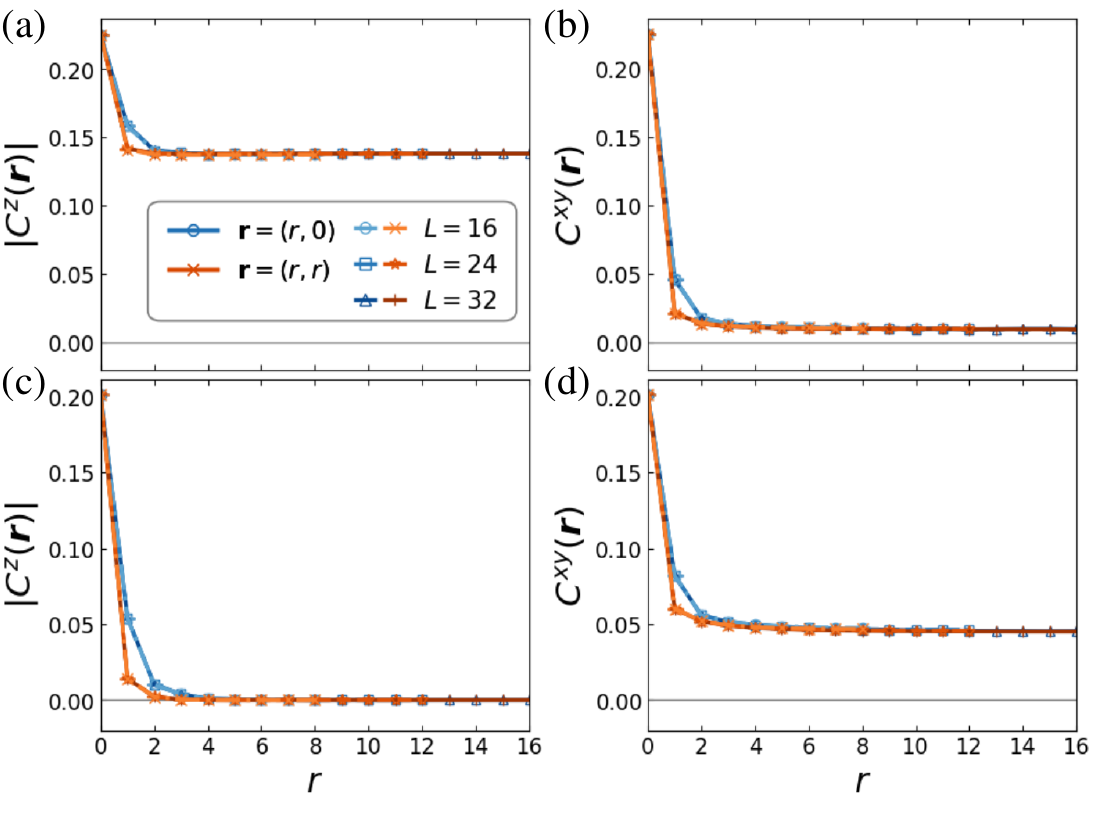}
    \caption{Diagonal spin correlations $|C^{z}(\boldsymbol{r})|$ and off-diagonal correlations $C^{xy}(\boldsymbol{r})$, evaluated along the lattice axial direction $\boldsymbol{r}=(r,0)$ and the diagonal direction $\boldsymbol{r}=(r,r)$, are shown in the DSS region (panels (a) and (b), $t=1.25$) and the DSF region (panels (c) and (d), $t=1.6$) for various system sizes ($L=\beta$). ($V=8$).}
    \label{fig:4}
\end{figure}

To directly validate the lattice symmetry breaking and the $U(1)$ symmetry breaking of the total magnetization, we further calculate the equal-time diagonal spin correlation functions $C^{z}(\boldsymbol{r}) = \frac{1}{N}\sum_{i} \langle \hat{S}_{i}^{z}\hat{S}_{i+\boldsymbol{r}}^{z}\rangle$ and the off-diagonal spin correlation functions $C^{xy}(\boldsymbol{r}) = \frac{1}{2N}\sum_{i} \langle \hat{S}_{i}^{x}\hat{S}_{i+\boldsymbol{r}}^{x}+\hat{S}_{i}^{y}\hat{S}_{i+\boldsymbol{r}}^{y} \rangle$ in the DSS phase (Figs.~\ref{fig:4}(a) and (b)) and in the DSF phase (Figs.~\ref{fig:4}(c) and (d)). The non-vanishing diagonal and off-diagonal spin correlations indicate that the spin part exhibits both the AFM out-of-plane and FM in-plane order in the DSS region, which can be experimentally verified by measuring the diagonal correlations $\langle \hat{S}^{z}_{i} \hat{S}^{z}_{j} \rangle$ and the in-plane correlations $\langle \hat{S}^{x}_{i} \hat{S}^{x}_{j}\rangle$ in the Rydberg tweezer arrays. The DSS here can be understood in the context of the defect mechanism proposed by Andreev-Lifshitz-Chester 
(ALC)~\cite{andreev1969quantum, PhysRevA.2.256}. In the ALC framework, supersolidity emerges when vacancies or defects (holes) coherently delocalize within an otherwise ordered crystal, originating from the competition between the kinetic energy gain through hole delocalization and the potential energy cost of distorting the crystalline order. Finite-size scaling analysis presented in the SM~\cite{supplemental}, suggests that the DSS-to-DSF transition is continuous. The nature of the AFM-to-DSS transition requires further investigation.

\begin{figure}[htb]
\includegraphics[width=0.99\linewidth]{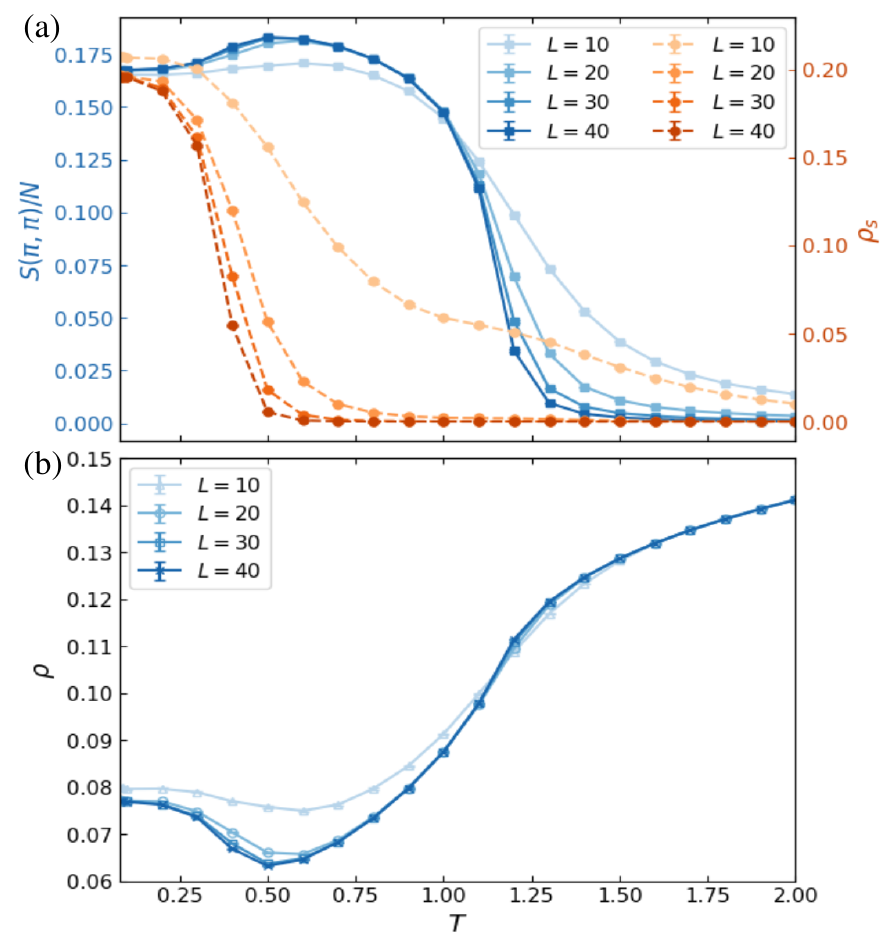}
\caption{(a) Structure factor $S(\pi,\pi)/N$ (blue solid lines) and  superfluid density $\rho_s$ (yellow dashed lines) versus temperature $T=1/\beta$ for various system sizes. 
(b) Hole density $\rho$ versus temperature $T$ is shown for the same four system sizes. ($V=8$ and $t=1.2$).}  \label{fig:fig5}
\end{figure}

{\it Finite-temperature transition.---} 
To investigate the thermal stability of the DSS phase, we conduct finite-temperature QMC simulations with parameters $V=8$ and $t=1.2$, which lies within the DSS region identified in Fig.~\ref{fig:fig3} in the ground state.
We employ the thermal annealing method~\cite{brooke1999quantum,Brooke_2001}, starting with high temperature in the disordered phase and then gradually decreasing temperature. 
The structure factor $S(\pi,\pi)/N$ and the superfluid density $\rho_s$ along the annealing path are shown in Fig.~\ref{fig:fig5}(a).
At low temperatures, both $S(\pi,\pi)/N$ and $\rho_s$ remain finite, confirming the thermal stability of the DSS phase. 
As the temperature increases, the superfluid density diminishes first, vanishing at approximately $T \simeq 0.5$. This transforms the DSS into a hole-doped AFM insulating phase with persistent diagonal order but zero superfluid density (The examination of the spin superfluidity is shown in the SM~\cite{supplemental}). As we increase the temperature further,  the AFM order eventually melts at $T \simeq 1.13$, yielding a completely disordered phase, which is consistent with the 2D Ising-type transition~\cite{supplemental}.

A remarkable feature of the DSS phase is the anomalous response of its spin structure factor to increasing temperature.  
In the temperature regime of  $T \lesssim 0.5$, the spin structure factor $S(\pi,\pi)/N$ increases anomalously with  temperature, instead of being suppressed by thermal fluctuations.
This counterintuitive phenomenon emerges from an intricate interplay between spin and hole degrees of freedom. 
From Fig.~\ref{fig:fig5}(b), the anomalous thermal response of $S(\pi,\pi)/N$ is tied with the non-monotonic behavior of the hole density $\rho$, which shows a minimum at $T \simeq 0.5$. 
This reduction in the hole density strengthens the spin-spin correlations. 
The hole density $\rho$ decrease originates from the loss of superfluidity, which localizes the holes and increases their effective mass to enforce a sparser distribution, thereby reducing their disruption to the spin background. 
This phenomenon, dubbed “thermal compensation ordering”, demonstrates how thermal fluctuations can actually enhance ordering in systems with multiple coupled degrees of freedom rather than universally destroying it.

{\it Experimental protocols.---} 
Realizing isotropic interactions in 2D entails aligning the magnetic field perpendicular to the atomic plane. While geometrically linked, the ratios $V/t$ and $J^{z,\perp}/t$ remain tunable via the lattice spacing $R$—leveraging the distinct power laws of dipolar exchange ($\propto R^{-3}$) and van der Waals interactions ($\propto R^{-6}$)—as well as through the specific choice of Rydberg states and atomic species (see SM~\cite{supplemental} for more details). Furthermore, highly independent control of the spin interactions is achievable via Floquet engineering~\cite{Geier_2021, PRXQuantum.3.020303} or microwave dressing~\cite{PhysRevA.84.033619}, effectively expanding the accessible parameter space in the Rydberg platform.

The Hamiltonian in Eq.~\eqref{eq:hamiltonian} conserves both the hole dopants $\hat{N}_{h}$ and the total magnetization $\hat{S}^{z}_{tot}$~\cite{qiao2025realization}. Although our grand-canonical QMC simulations incorporate fluctuations in these quantities, experiments typically operate in fixed-sector settings. A feasible protocol is as follows: (i) Prepare a low-energy initial state with target $N_{h}$ and $S^{z}_{tot}$ using site-resolved light shifts; (ii) quench the light shifts~\cite{chen2023continuous, qiao2025realization} to initiate evolution under $\hat{H}_{tJV}$; and (iii) extract diagonal and off-diagonal correlations via projective measurements to identify the corresponding phases after thermalization~\footnote{The relaxation to equilibrium is expected due to the non-integrable nature of the interacting $t$-$J$-$V$ Hamiltonian. Both the two-dimensional geometry and the long-range interactions promote quantum chaos, which, according to the Eigenstate Thermalization Hypothesis (ETH), allows the local observables to relax to steady-state values.}.

{\it Conclusion.---} 
In summary, we have investigated the bosonic $t$–$J$–$V$ model on the square lattice using large-scale QMC simulations, motivated by its recent realization in Rydberg atom arrays~\cite{qiao2025realization}. We identified a rich ground-state phase diagram comprising both conventional AFM and DSF phases, as well as an exotic DSS phase. The DSS phase arises from the interplay between spin and hole degrees of freedom under long-range interactions, and is characterized by the simultaneous breaking of the lattice symmetry and two $U(1)$ symmetries, appearing above a critical hole–hole repulsion. Our finite-temperature analysis further reveals a two-step melting process together with a counterintuitive thermal enhancement of crystalline order. These findings provide direct theoretical guidance for ongoing experimental efforts on this model. 
The investigation of these strongly correlated quantum phases in the bosonic $t$–$J$–$V$ model is expected to shed new light on the physics of doped quantum magnetism~\cite{Bohrdt_2021,PhysRevLett.132.230401}.

{\it Acknowledgments}.--- 
We acknowledge helpful discussion with Yan-Cheng Wang, Lukas Homeier, Dong-Xu Liu, Heng Shen and Xue-Feng Zhang.  
This work is supported by  the Innovation Program for Quantum Science and Technology of China (Grant No. 2024ZD0300100),
the National Basic Research Program of China (Grants No. 2021YFA1400900), 
Shanghai Municipal Science and Technology (Grant No. 25TQ003, 2019SHZDZX01, 24DP2600100). 
ZY thanks the Scientific Research Fund for Distinguished Young Scholars of the Education Department of Anhui Province (No.2022AH020008) and the Scientific Research Project of Westlake University (No. WU2024B027). The authors also acknowledge the HPC Centre of Westlake University and Beijing PARATERA Tech Co., Ltd. for providing HPC resources.

 \bibliography{references}

\newpage~
\newpage~
%\appendix

\setcounter{equation}{0}
\setcounter{figure}{0}
\setcounter{table}{0}
\setcounter{section}{0}
\setcounter{page}{1}
\renewcommand{\theequation}{S\arabic{equation}}
\renewcommand{\thefigure}{S\arabic{figure}}
\renewcommand{\thetable}{S\Roman{table}}

%%%%%%%%%%%%%%%%%%%%%%%%%%%%%%%%%%%%%%%%%%%%%%%%%%%%%
\onecolumngrid

\begin{center} \huge{Supplemental Materials: Double Supersolid Phase in a Bosonic $t$-$J$-$V$ Model with Rydberg Atoms\\}
\end{center}

%\maketitle

\vspace*{2cm}

%{\color{red} Why is overleaf not showing the title of the supplemental??}
In this Supplemental Material, we describe the SSE-QMC algorithm for the $t$-$J$-$V$ model together with additional results that support the claims made in the main text.

%%%%%%%%%%%%%%%%%%%%%%%%%%%%%%%%%%%%%%%%%%%%%%%%%%%%%
\section{Quantum Monte Carlo Scheme}
\label{app:QMC}
%%%%%%%%%%%%%%%%%%%%%%%%%%%%%%%%%%%%%%%%%%%%%%%%%%%%%
The Hamiltonian of the $t$-$J$-$V$ model is 
\begin{equation}
    \hat{H}_{tJV} = -\sum_{i<j}\sum_{\sigma=\downarrow,\uparrow}\frac{t}{r_{ij}^3}\left(\hat{a}_{i,\sigma}^\dagger\hat{a}_{j,h}^\dagger\hat{a}_{i,h}\hat{a}_{j,\sigma}+\mathrm{h.c.}\right) 
    + \sum_{i<j}\frac{1}{r_{ij}^{6}}\left[J^{z}\hat{S}_{i}^{z}\hat{S}_{j}^{z}+\frac{J^{\perp}}{2}\left(\hat{S}_{i}^{+}\hat{S}_{j}^{-}+\mathrm{h.c.}\right)\right] 
    + \sum_{i<j}\frac{V}{r_{ij}^{6}}\hat{n}_{i}^{h}\hat{n}_{j}^{h}. 
\end{equation}
Here we set $\mu=0$. We decompose this Hamiltonian into elementary operators defined on each bond:
\begin{equation}
    \begin{aligned}
        \hat{H}_{0,0} &= \mathbb{I},\\
        \hat{H}_{1,b} &=  \sum_{\sigma=\downarrow,\uparrow}\frac{t}{r_{ij}^3} \left(\hat{a}_{i,\sigma}^\dagger\hat{a}_{j,h}^\dagger\hat{a}_{i,h}\hat{a}_{j,\sigma}+\mathrm{h.c.}\right), \\
        \hat{H}_{2,b} &= -\frac{J^{\perp}}{2 r_{ij}^6} \left(\hat{S}_{i}^{+}\hat{S}_{j}^{-}+\mathrm{h.c.}\right), \\
        \hat{H}_{3,b} &= - \frac{J^{z}}{r_{ij}^6}  \hat{S}_{i}^{z}\hat{S}_{j}^{z} - \frac{V}{r_{ij}^{6}}\hat{n}_{i}^{h}\hat{n}_{j}^{h} + C_{ij},
    \end{aligned}
\end{equation}
where the $r^{-6}$ terms are truncated at  $r_c=2$ and the $r^{-3}$ terms at $r_c=4$ (in units of the lattice spacing). The constant $C_{ij}$ is added to ensure all matrix elements in $\hat{H}_{3,b}$ are non-negative:
\begin{equation}
    C_{ij} = \max\Big(|\min(0,-V_{ij},-|J^{z}_{ij}|/4)| + \epsilon|\min(0,-V_{ij},-|J^{z}_{ij}|/4)|, t_{ij}\Big),
\end{equation}
with $V_{ij}= V/{r_{ij}^{6}}$, $J^{z,\perp}_{ij}= J^{z,\perp}/{r_{ij}^{6}}$ and $t_{ij}=t/{r_{ij}^{3}}$. $\epsilon=0.1$ is a fixed multiplicative constant during simulations.  

Now $\hat{H}_{tJV}=-\sum_{i=1}^{3} \sum_{b} \hat{H}_{i,b}$ and the partition function $Z=\operatorname{Tr} e^{-\beta H}$ can be expressed as a power series expansion:
\begin{equation}
    Z=\sum_\alpha\sum_{S_M}\frac{\beta^n(M-n)!}{M!}\langle\alpha|\prod_{i=1}^M \hat{H}_{a_i,b_i}|\alpha\rangle,
    \label{eq:partition}
\end{equation}
where $M$ is the cutoff of the expansion series and $S_{M}$ represents a particular sequence of $n$ non-identity elementary operators in imaginary time. The matrix elements in Eq.~\eqref{eq:partition} are constructed from matrix elements in the spin $\sigma^{z}$ and hole occupation basis:
\begin{equation}
    \begin{aligned}
        \bra{\uparrow h}\hat{H}_{1,b}\ket{h\uparrow}&=\bra{h \uparrow }\hat{H}_{1,b}\ket{\uparrow h} = \bra{\downarrow h}\hat{H}_{1,b}\ket{h\downarrow} = \bra{ h \downarrow}\hat{H}_{1,b}\ket{\downarrow h} = t_{ij},  \\
        \bra{\uparrow \downarrow}\hat{H}_{2,b}\ket{\downarrow \uparrow} &=
        \bra{\downarrow \uparrow}\hat{H}_{2,b}\ket{\uparrow \downarrow} = - J^{\perp}_{ij}/2, \\
        W^{(1)}_{ij} &=\bra{\uparrow \downarrow}\hat{H}_{3,b}\ket{ \uparrow \downarrow}=\bra{\downarrow \uparrow}\hat{H}_{3,b}\ket{ \downarrow \uparrow} = C_{ij} + J^{z}_{ij}/4, \\
        W^{(2)}_{ij} &=\bra{\uparrow \uparrow}\hat{H}_{3,b}\ket{ \uparrow \uparrow}=\bra{\downarrow \downarrow}\hat{H}_{3,b}\ket{ \downarrow \downarrow} = C_{ij} - J^{z}_{ij}/4,  \\  
        W^{(3)}_{ij} &=\bra{\uparrow h}\hat{H}_{3,b}\ket{\uparrow h}=\bra{h\uparrow}\hat{H}_{3,b}\ket{h\uparrow}= \bra{\downarrow h}\hat{H}_{3,b}\ket{\downarrow h} = \bra{h\downarrow }\hat{H}_{3,b}\ket{h\downarrow} = C_{ij},\\
        W^{(4)}_{ij} &=\bra{h h}\hat{H}_{3,b}\ket{hh} = C_{ij} - V_{ij},
    \end{aligned}
\end{equation}
where the subscripts $i,j$ in $W^{(1,2,3,4)}$ indicate spatial dependence. Since all the matrix elements carry strictly positive weights in the chosen parameter regime ($J^{z} >0$, $J^{\perp}<0$, $t>0$), the model is sign-problem-free and can thus be tackled with Quantum Monte Carlo techniques.

The updating scheme contains two major updates: the diagonal update and the directed-loop update, described as follows.

(1) Diagonal update:  
The diagonal update proceeds by visiting every imaginary time slice $p \in \{1,2,\dots,M\}$ and either removing or inserting a diagonal operator according to the following procedure~\cite{merali2024stochastic}.  

\quad (a) For a diagonal operator $\hat{H}_{3,b}$ encountered, remove it ($n\rightarrow n-1$) with probability
\begin{equation}
    P(n\rightarrow n-1) = \min\left(\frac{M-n+1}{\beta \mathcal{N}}, 1\right),
\end{equation}
with $\mathcal{N}=\sum_{b} \max(W^{(1)}_{ij}, W^{(2)}_{ij}, W^{(3)}_{ij}, W^{(4)}_{ij})$ the normalizing factor.

\quad (b) For an identity operator $\hat{H}_{0,0}$ encountered, decide whether to insert ($n\rightarrow n+1$) the diagonal operator $\hat{H}_{3,b}$ with probability
\begin{equation}
     P(n\rightarrow n+1) = \min\left(\frac{\beta \mathcal{N}}{M-n},1\right).
\end{equation}

\quad (c) If insertion is accepted, use the Alias method~\cite{walker1974newfastmethod, walker1977anefficientmethod,vose1991linear} to sample over spatial bonds according to the distribution $P_{b} = \max(W^{(1)}_{ij}, W^{(2)}_{ij}, W^{(3)}_{ij}, W^{(4)}_{ij}) / \mathcal{N}$.  

\quad (d) After selecting bond $b$, read the state at the current imaginary time slice $p$ to evaluate the actual weight $W^{\text{actual}}_{ij} \in \{W^{(1)}_{ij}, W^{(2)}_{ij}, W^{(3)}_{ij}, W^{(4)}_{ij}\}$. Insert the operator $\hat{H}_{3,b}$ at bond $b$ with probability
\begin{equation}
    P = \frac{W^{\text{actual}}_{ij}}{ \max(W^{(1)}_{ij}, W^{(2)}_{ij}, W^{(3)}_{ij}, W^{(4)}_{ij})}.
\end{equation}

\quad (e) For an off-diagonal operator $\hat{H}_{1,b}$ or $\hat{H}_{2,b}$ encountered, propagate the state $\ket{\alpha_{p}} \propto \hat{H}_{1,b/2,b} \ket{\alpha_{p-1}}$.

\quad (f) Repeat the procedure for the next imaginary time slice.

(2) Directed-loop update:
The directed-loop update begins by randomly selecting one of the vertex legs as an initial entrance leg. The state ($\ket{h},\ket{\uparrow},\ket{\downarrow}$) of the entrance leg determines the subsequent steps.  

\quad (a) If the state is $\ket{\uparrow}$ or $\ket{\downarrow}$, either flip the spin or annihilate it to create a hole with probability $1/2$ (corresponding to adding a $\pm 1$ or $\pm 2$ operator at the entrance leg).  

\quad (b) If the state is $\ket{h}$, either change the hole to $\ket{\uparrow}$ or to $\ket{\downarrow}$ with probability $1/2$ (corresponding to adding a $\pm 1$ operator at the entrance leg).  

\quad (c) Choose the exit leg with probability determined by pre-computed weights. These probabilities are determined from matrix elements obtained by updating both the entrance and exit legs. For the detailed construction see Ref.~\cite{PhysRevLett.105.120603}.  

\quad (d) The next entrance leg is the one connected to the exit leg. The process continues until a closed loop is formed.

\begin{figure}[htb]
    \centering
    \includegraphics[width=0.95\linewidth]{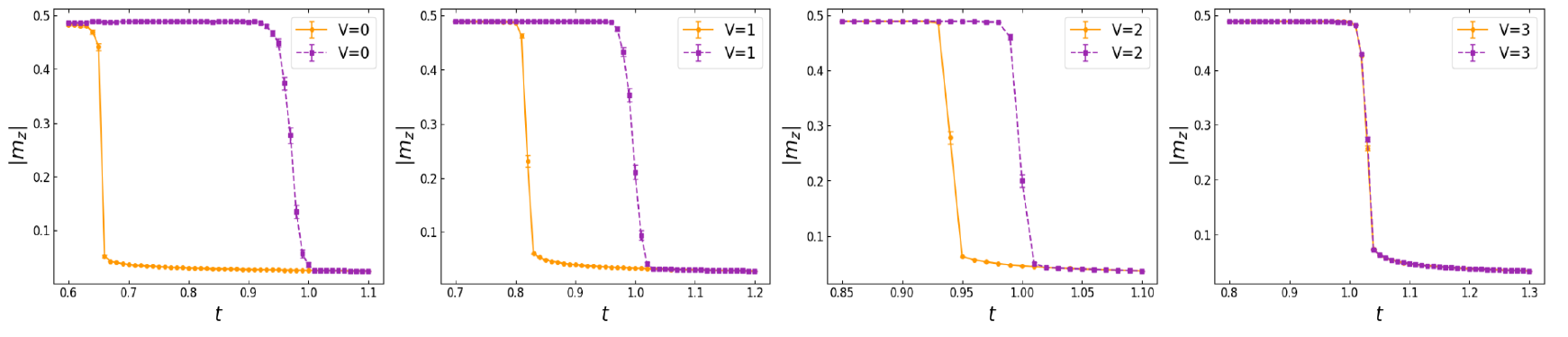}
    \caption{Hysteresis of the Néel order parameter $|m_z|$ during QMC simulations with different $V$. The yellow solid line shows results from forward annealing, while the purple dashed line shows results from backward annealing. ($L=\beta=16$).}
    \label{fig:S1}
\end{figure}

During our simulations, we found that conventional QMC methods struggled to efficiently sample the small-$t$ region. To overcome this challenge, we employed the annealing method~\cite{aranson2001dynamics,mitra2018quantumquenchdynamics,yan2023emergent}, where the system was initialized at large $t$ and $t$ was gradually decreased through successive simulations. After completing the forward annealing process, we also performed the reverse procedure by gradually increasing $t$. This approach revealed a hysteresis loop indicative of a first-order phase transition, as shown in Fig.~\ref{fig:S1}. 

Notably, as $V$ increases, the hysteresis loop progressively shrinks, signaling a weakening of the first-order character. At $V=3$, the forward and backward results nearly converge, suggesting the proximity of a tricritical point.

%%%%%%%%%%%%%%%%%%%%%%%%%%%%%%%%%%%%%%%%%%%%%%%%%%%%%
%%%%%%%%%%%%%%%%%%%%%%%%%%%%%%%%%%%%%%%%%%%%%%%%%%%%%
\section{Finite-Size Scaling Analysis}
\label{app:FSS}
\begin{figure}[htb]
    \centering
    \includegraphics[width=0.48\linewidth]{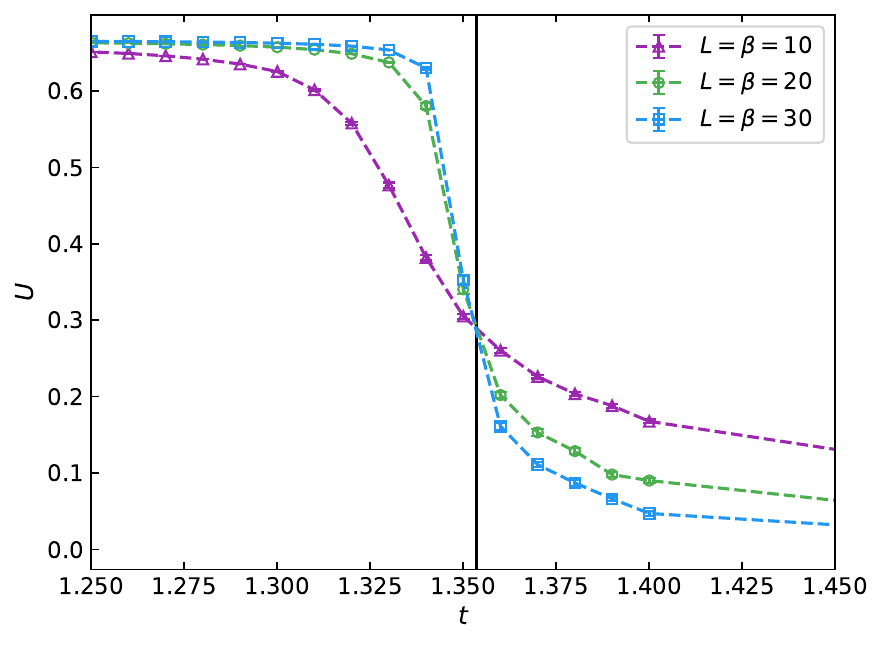}
    \caption{Binder cumulant $U$ as a function of $t$ for different sizes with $V=8$. The black solid line marks $t_c=1.354(1)$.}
    \label{fig:S2}
\end{figure}

\begin{figure}[htb]
    \centering
    \includegraphics[width=0.48\linewidth]{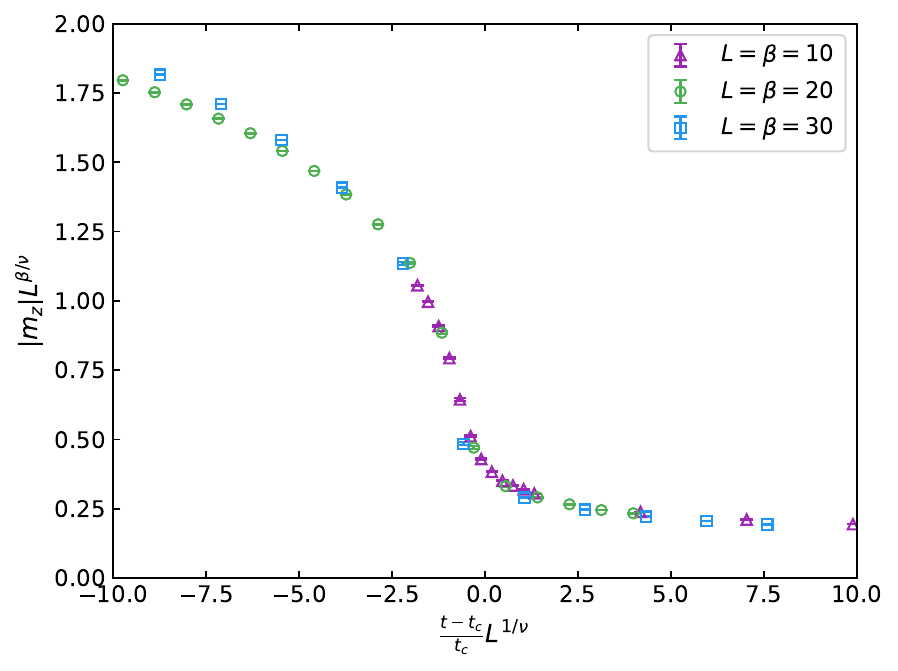}
    \caption{Data collapse of the Néel order parameter $|m_z|$ near the critical point $t_c$ with  $\nu=0.6301$ and $\beta=0.3265$.}
    \label{fig:S3}
\end{figure}

%%%%%%%%%%%%%%%%%%%%%%%%%%%%%%%%%%%%%%%%%%%%%%%%%%%%%
We perform finite-size scaling analysis of our simulation results. We first examine the phase transition from the DSS phase to the DSF phase. We consider the Binder cumulant, which is a dimensionless quantity to locate the quantum critical point,
\begin{equation}
    U=1-\frac{\langle m^4\rangle}{3\langle m^2\rangle^2},
\end{equation}
with $m = \frac{1}{N} \sum_{j}  e^{i\boldsymbol{Q}\cdot\mathbf{r}_j} S^{z}_{j}$ and $\boldsymbol{Q}=(\pi,\pi)$. The results are shown in Fig.~\ref{fig:S2}, with the crossing point located at $t_c = 1.354(1)$. We further collapse the Néel order parameter $|m_z|$ using the determined $t_c$ according to the finite-size scaling relation 
\[
    |m_z|\cdot L^{\beta/\nu}=f\left[\frac{t-t_c}{t_c}\cdot L^{1/\nu}\right],
\]
as shown in Fig.~\ref{fig:S3}. The critical exponents $\nu=0.6301$ and  $\beta=0.3265$ used for the data collapse belong to the (2+1)D Ising universality class. This is expected as the van der Waals interaction that stabilizes the AFM order decays rapidly and is effectively short-ranged.

\begin{figure}[htb]
    \centering
    \includegraphics[width=0.5\linewidth]{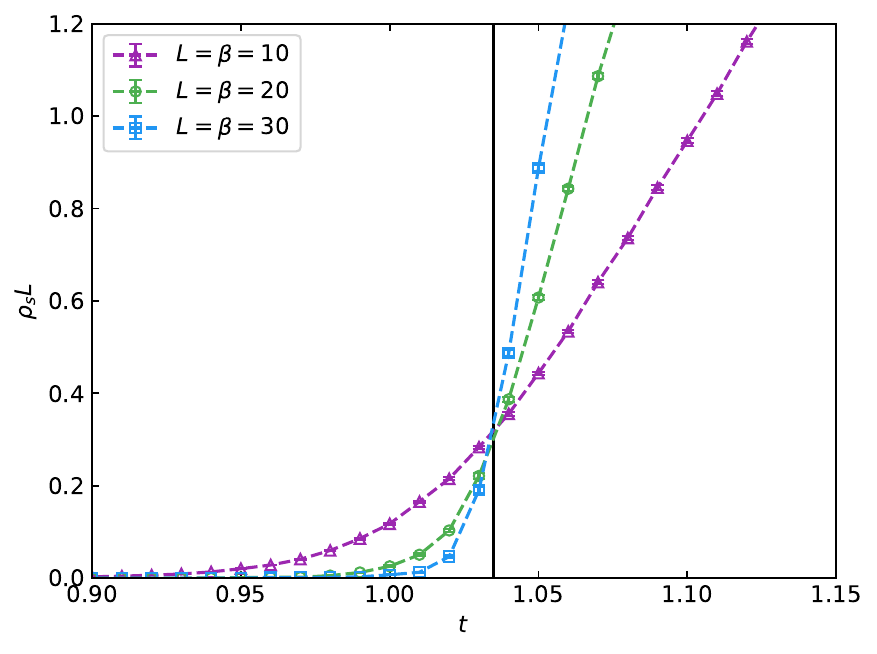}
    \caption{Scaled superfluid density $\rho_s L$ for various system sizes with $V=8$. The black solid line marks $t_c=1.035(1)$.}
    \label{fig:S4}
\end{figure}

We then consider the phase transition from the AFM phase to the DSS phase, which breaks the $U(1)$ symmetry.  In Fig.~\ref{fig:S4}, we plot the scaled superfluid density $\rho_s L$  for different system sizes (assuming $z=1$), which intersect at $t_c=1.035(1)$.

However, we note that due to the truncation of long-range interactions and the interplay between two $U(1)$ symmetries (spin and hole), the current data do not provide conclusive evidence that the AFM-to-DSS transition is continuous. We leave this question for future investigation.

\begin{figure}[htb]
    \centering
    \includegraphics[width=0.5\linewidth]{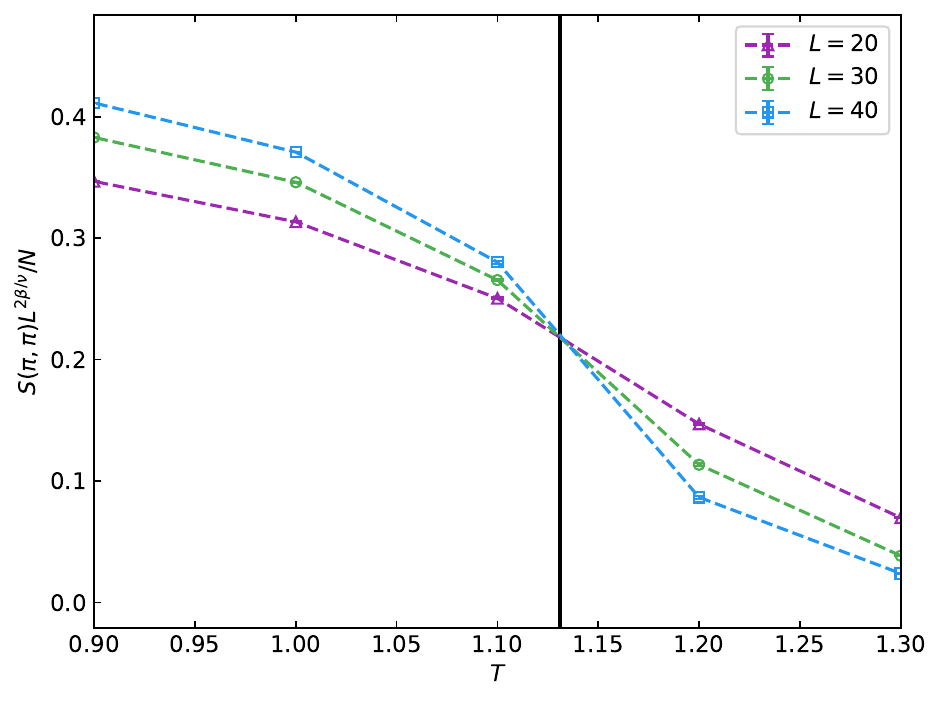}
   \caption{Finite-size scaling of the reduced structure factor $S(\pi,\pi)/N$. $S(\pi,\pi)L^{2\beta/\nu}/N$ is plotted for various system sizes near the critical point $T_c$, with $2\beta/\nu=0.25$. The black solid line marks $T_c=1.131(2)$. ($V=8$ and $t=1.2$).}
    \label{fig:S5}
\end{figure}

Finally, we consider the finite-temperature transition that melts the AFM phase into the disordered phase. In Fig.~\ref{fig:S5}, we plot the scaled reduced structure factor $S(\pi,\pi)L^{2\beta/\nu}/N$ for different system sizes (excluding $L=10$ due to strong finite-size effects).  The universal crossing point identifies a critical temperature $T_{c}=1.131(2)$, and the scaling exponents are consistent with the 2D Ising universality class ($2\beta/\nu = 0.25$).

%%%%%%%%%%%%%%%%%%%%%%%%%%%%%%%%%%%%%%%%%%%%%%%%%%%%%
%%%%%%%%%%%%%%%%%%%%%%%%%%%%%%%%%%%%%%%%%%%%%%%%%%%%%
\section{Extrapolation of spin correlation}
\label{app:Extra}
\begin{figure}[htb]
    \centering
    \includegraphics[width=0.6\linewidth]{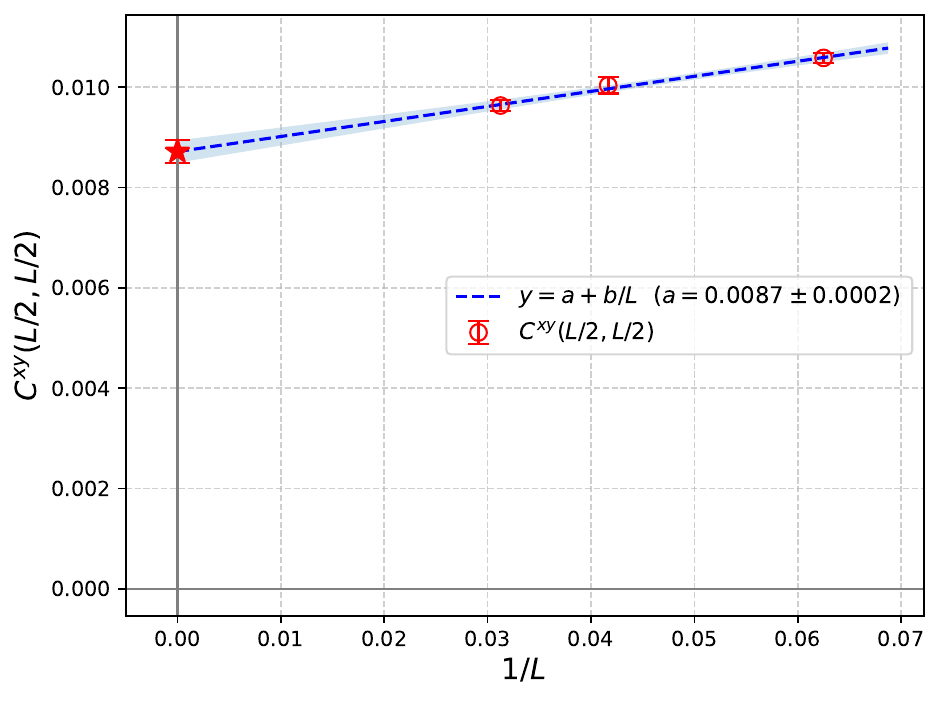}
    \caption{Finite-size scaling of the off-diagonal spin correlation $C^{xy}(\boldsymbol{r})$ with $\boldsymbol{r} = (L/2,L/2)$ versus the inverse system size $1/L$ for $L=16, 24, 32$. The dashed line represents a linear fit, yielding an extrapolated intercept $a=0.0087(2)$ in the thermodynamic limit ($1/L\to 0$). ($V=8$ and $t=1.25$).}
    \label{fig:S6}
\end{figure}

%%%%%%%%%%%%%%%%%%%%%%%%%%%%%%%%%%%%%%%%%%%%%%%%%%%%%
To corroborate the existence of off-diagonal long-range order (LRO) in the thermodynamic limit, we performed a finite-size extrapolation of the off-diagonal spin correlation function. We examine the value of $C^{xy}(\boldsymbol{r})$ at the largest possible separation $\boldsymbol{r} = (L/2,L/2)$ for different system sizes $L$. Fig.~\ref{fig:S6} shows the scaling of $C^{xy}(L/2,L/2)$ versus the inverse system size $1/L$ in the DSS region. The data points are well-fitted by a linear function $f(x)=a+bx$. The extrapolation to the thermodynamic limit ($1/L \to 0$) yields a finite intercept $a=0.0087(2)$, providing direct evidence for true off-diagonal LRO in the DSS phase.

%%%%%%%%%%%%%%%%%%%%%%%%%%%%%%%%%%%%%%%%%%%%%%%%%%%%%
\section{Superfluid density of spin part}
\begin{figure}[htb]
    \centering
   \includegraphics[width=0.99\linewidth]{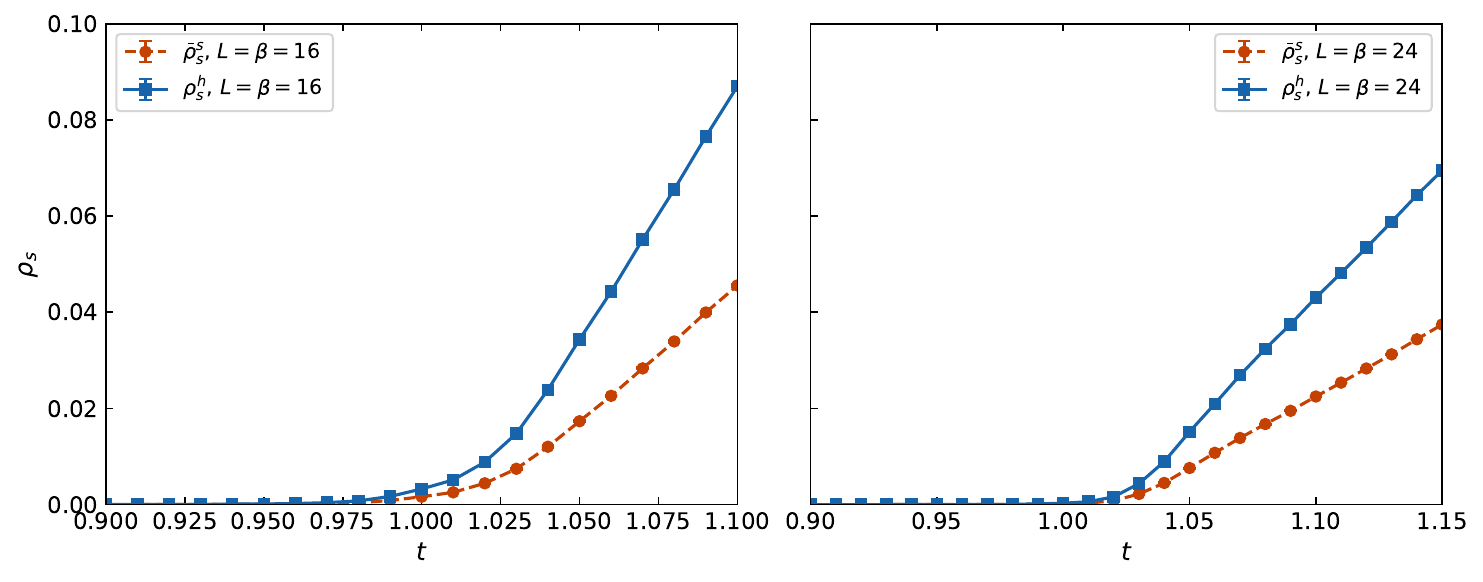}
    \caption{Superfluid density as a function of $t$ for spin and hole components. The left panel shows the $L=16$ result, while the right panel shows the $L=24$ result. ($V=8$).}
    \label{fig:S7}
\end{figure}

We examine the superfluid density associated with the spin component, following the similar definition as for the hole part
\begin{equation}
    \rho_{s}^{\uparrow/\downarrow}=\frac{L^{2-d}\langle W_{x,\uparrow/\downarrow}^{2} + W_{y,\uparrow/\downarrow}^{2}\rangle}{2d\beta t},
\end{equation}
where $W_{x/y,\uparrow/\downarrow}$ refers to the winding numbers of the spin up/down particles in the $x/y$-directions. Considering the symmetry between spin up and spin down, we plot the average spin-component superfluid density $\bar{\rho}_{s}^{s} = \tfrac{1}{2}(\rho_{s}^{\uparrow} + \rho_{s}^{\downarrow})$.
For the AFM-to-DSS transition the result is shown in Fig.~\ref{fig:S7}. 

\begin{figure}[htb]
    \centering
    \includegraphics[width=0.5\linewidth]{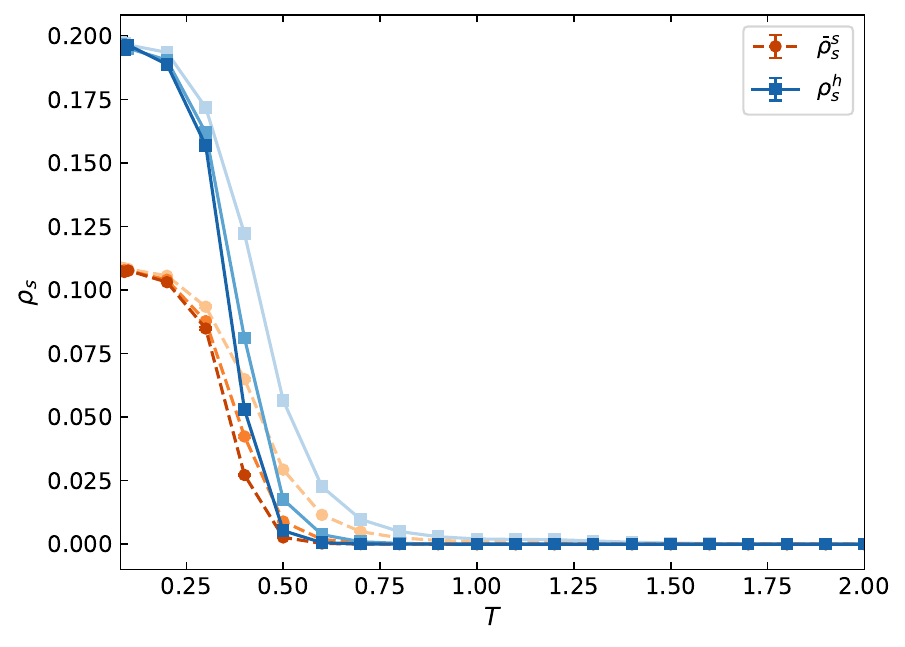}
    \caption{Superfluid density as a function of temperature $T$ for spin and hole components. The color gradient from light to dark indicates increasing system sizes ($L=20, 30, 40$). ($V=8$ and $t=1.2$).}
    \label{fig:S8}
\end{figure}

To confirm that spin superfluidity also vanishes during the thermal phase transition, we present measurements in Fig.~\ref{fig:S8}, which demonstrate that the spin superfluid density behaves similarly to the hole superfluid density, consistently diminishing around the same transition temperature.

%%%%%%%%%%%%%%%%%%%%%%%%%%%%%%%%%%%%%%%%%%%%%%%%%%%%%
%%%%%%%%%%%%%%%%%%%%%%%%%%%%%%%%%%%%%%%%%%%%%%%%%%%%%
\section{Hole density and correlations}

\begin{figure}[htb]
    \centering
    \includegraphics[width=0.5\linewidth]{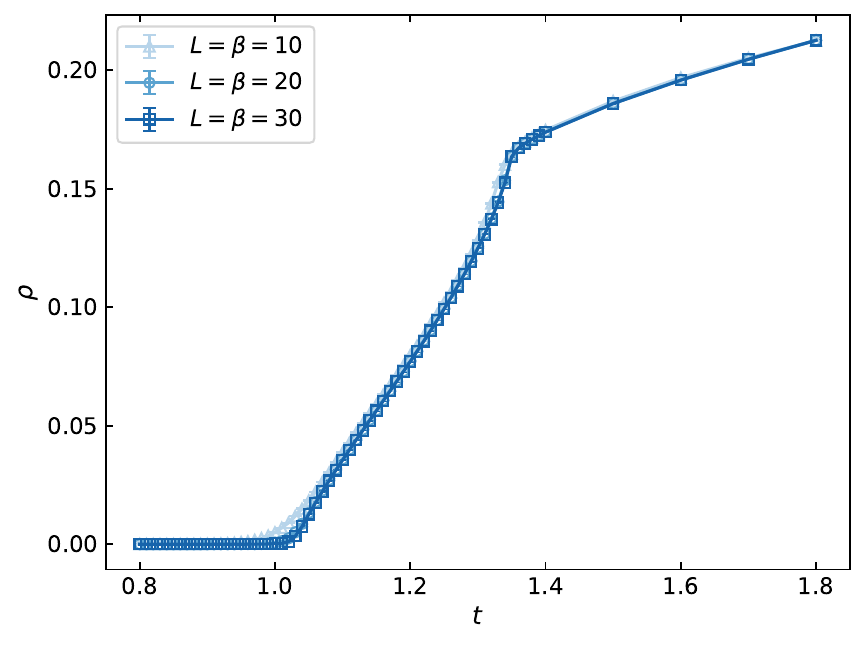}
    \caption{Hole density $\rho$ as a function of the tunneling strength $t$ for various system sizes. ($V=8$).}
    \label{fig:S9}
\end{figure}

In Fig.~\ref{fig:S9}, we plot the hole density $\rho=\frac{1}{N} \langle \sum_{j}  \hat{n}^{h}_{j}\rangle$ for the same parameters as in Fig.3 in the main text. The hole density $\rho$ vanishes deep inside the small-$t$ region, which leads to an effective pure spin-$1/2$ model with
$\hat{H}_{J}=\sum_{i<j}\frac{1}{r_{ij}^{6}}\left[J^{z}\hat{S}_{i}^{z}\hat{S}_{j}^{z}+\frac{J^{\perp}}{2}\left(\hat{S}_{i}^{+}\hat{S}_{j}^{-}+\mathrm{h.c.}\right)\right].$

\begin{figure}[htb]
    \centering
    \includegraphics[width=0.99\linewidth]{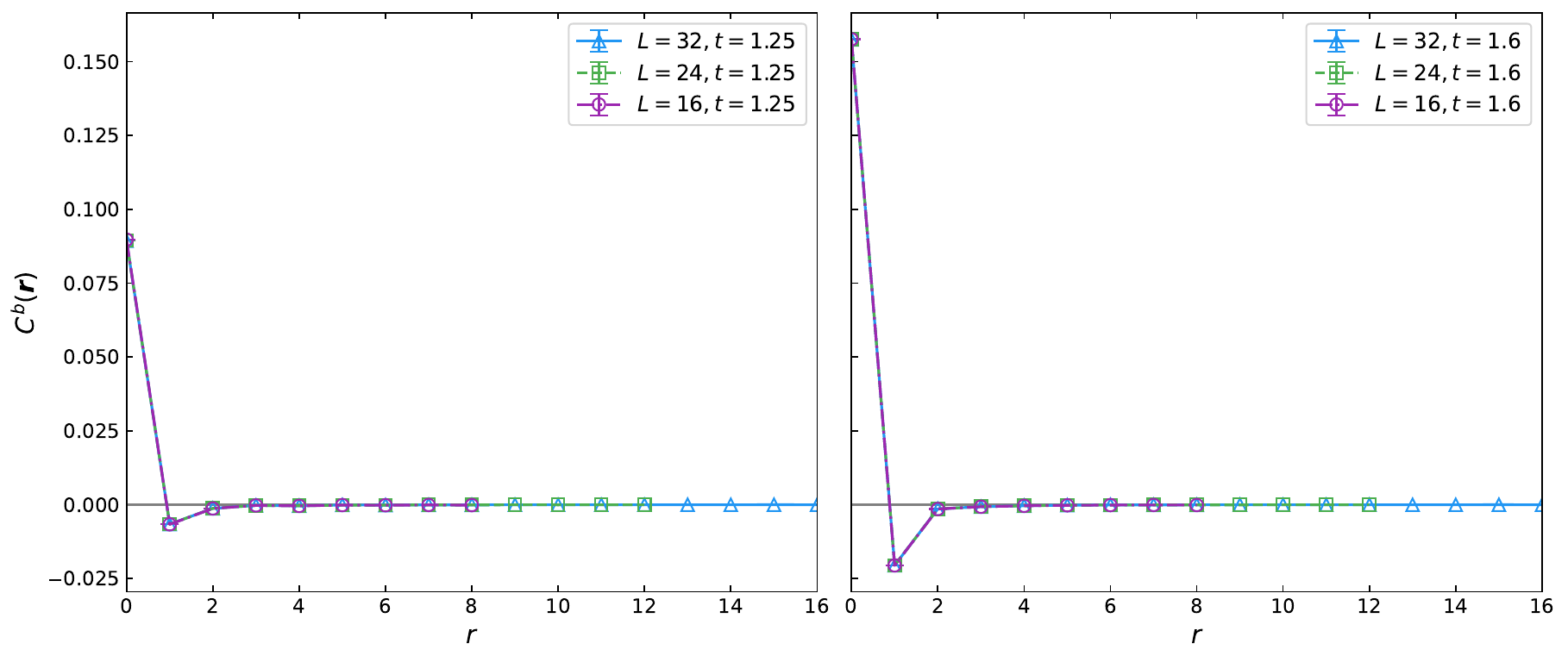}
    \caption{Density–density correlations of bosonic holes, $C^{b}(\boldsymbol{r})$, along the $x$-direction, $\boldsymbol{r} = (r,0)$. Results are shown for various system sizes ($L=\beta$) in the DSS region (left panel, $t=1.25$) and in the DSF region (right panel, $t=1.6$). ($V=8$).}
    \label{fig:S10}
\end{figure}

In Fig.~\ref{fig:S10}, we plot the density–density correlations of bosonic holes:
\begin{equation}
    C^{b}(\boldsymbol{r}) = \frac{1}{N}\sum_{i} \left[ \langle \hat{n}_{i}^{h} \hat{n}_{i+\boldsymbol{r}}^{h}\rangle - \langle \hat{n}_{i}^{h}\rangle  \langle \hat{n}_{i+\boldsymbol{r}}^{h}\rangle \right],
\end{equation}
along the $x$-direction in the DSS and DSF regions. Except at the nearest-neighbor point—where a pronounced blockade effect dominates due to the strong hole-hole repulsion— $C^{b}(\boldsymbol{r})$ decays rapidly to zero, consistent with a uniform density profile.

%%%%%%%%%%%%%%%%%%%%%%%%%%%%%%%%%%%%%%%%%%%%%%%%%%%%%
\section{Comparison between different truncation ranges and boundary conditions}
\label{app:Boundary}
\begin{figure}[htb]
    \centering
    \includegraphics[width=0.99\linewidth]{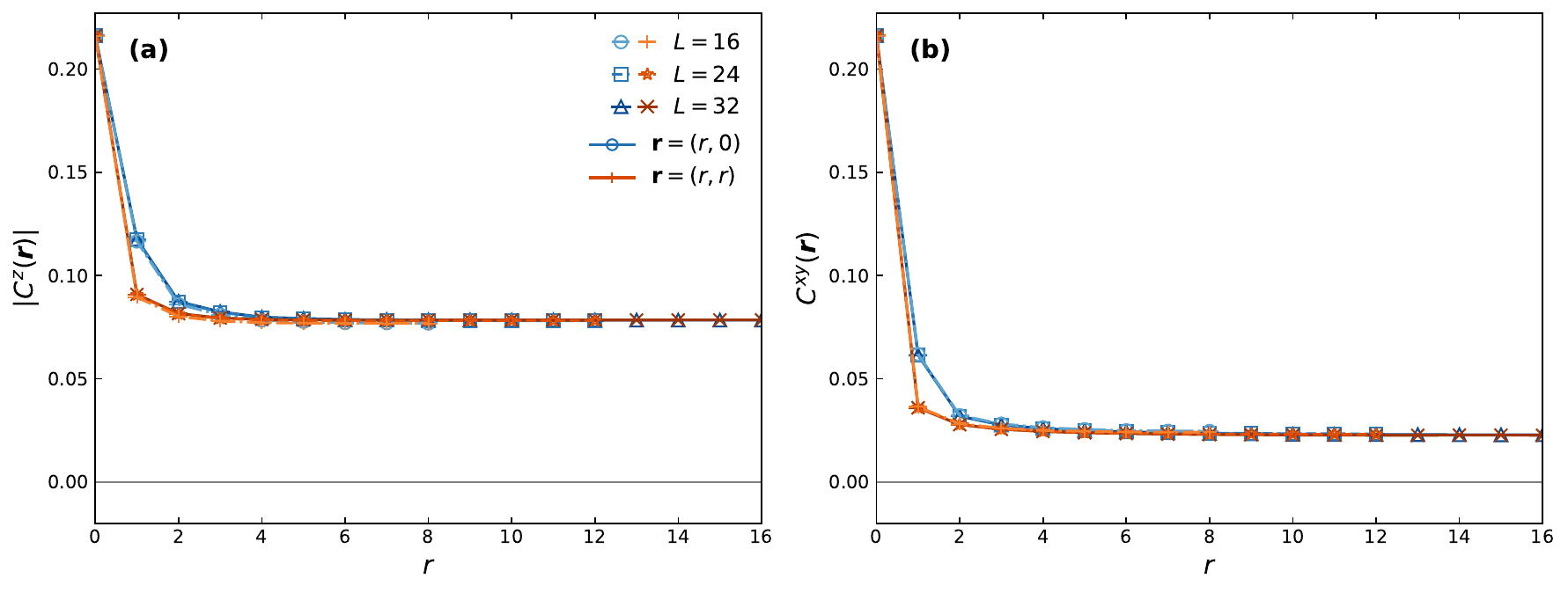}
    \caption{Diagonal spin correlations $|C^{z}(\boldsymbol{r})|$ and off-diagonal spin correlations $C^{xy}(\boldsymbol{r})$ with $\boldsymbol{r} = (r,0)$ and $\boldsymbol{r} = (r,r)$, are shown in the DSS region for various system sizes ($L=\beta$). ($V=8$,  $t=1.25$, with dipolar interactions truncated at $r_c=6$ in lattice units).}
    \label{fig:S11}
\end{figure}

%%%%%%%%%%%%%%%%%%%%%%%%%%%%%%%%%%%%%%%%%%%%%%%%%%%%%
Complementing the main text, we present additional simulations performed with an extended dipolar cutoff of $r_c = 6$. Fig.~\ref{fig:S11} displays the spin correlations within the DSS regime using this extended range. The DSS phase persists under this extension, confirming that $r_c = 4$ is sufficient to capture the essential long-range physics stabilizing the phase.

\begin{figure}[htb]
    \centering
    \includegraphics[width=0.99\linewidth]{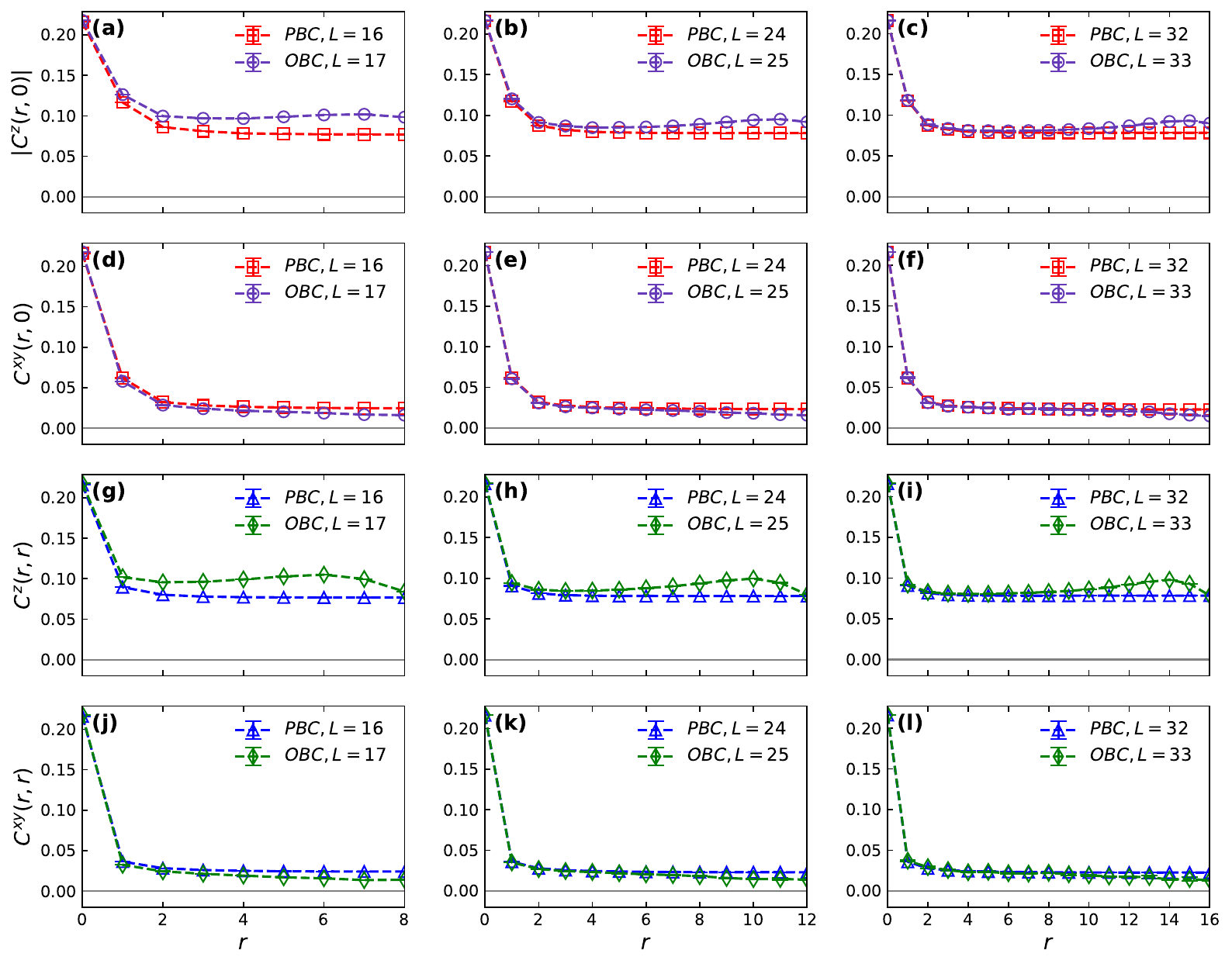}
    \caption{Comparison of the diagonal/off-diagonal spin correlations under different boundary conditions in the DSS region, shown for various system sizes ($L=\beta$). Panels (a–c) display $|C^{z}(\mathbf{r})|$ along the $\mathbf{r}=(r,0)$ direction, while panels (d–f) show $C^{xy}(\mathbf{r})$ along $\mathbf{r}=(r,0)$. Panels (g–i) present $C^{z}(\mathbf{r})$ along the diagonal $\mathbf{r}=(r,r)$, and panels (j–l) show the corresponding $C^{xy}(\mathbf{r})$. ($V=8$,  $t=1.25$, with dipolar interactions truncated at $r_c=6$ in lattice units).}
    \label{fig:S12}
\end{figure}

To examine whether the DSS phase is sensitive to boundary conditions, we performed additional simulations under open boundary conditions (OBC), which more closely reflect realistic experimental settings. Specifically, we compare the PBC results for the diagonal and off-diagonal correlations, $C^{z}(\boldsymbol{r})$ and $C^{xy}(\boldsymbol{r})$, along $\boldsymbol{r}=(r,0)$ and $\boldsymbol{r}=(r,r)$ for system sizes $L=16,24,32$, with the corresponding OBC results at $L=17,25,33$. Under OBC, translation symmetry is absent, so correlations are measured from the central site without spatial averaging; the system size $L$ is therefore chosen to be odd to ensure a unique center. This setup enables a direct and faithful comparison between PBC and OBC, as shown in Fig.~\ref{fig:S12}. At finite system sizes, OBC results exhibit boundary-induced features, including oscillations in $C^{z}(\boldsymbol{r})$ and enhanced decay in $C^{xy}(\boldsymbol{r})$ near the edges. However, these boundary effects tend to diminish systematically as we increase the system size,  and the OBC results within the bulk converge toward their PBC counterparts. This confirms that the choice of boundary conditions does not affect the essential bulk physics of the corresponding phases once the system is sufficiently large.

%%%%%%%%%%%%%%%%%%%%%%%%%%%%%%%%%%%%%%%%%%%%%%%%%%%%%
\section{Rydberg tweezer implementation}
\label{app:Rydberg}
%%%%%%%%%%%%%%%%%%%%%%%%%%%%%%%%%%%%%%%%%%%%%%%%%%%%%
\begin{figure}[htbp]
    \centering
    \includegraphics[width=0.6\linewidth]{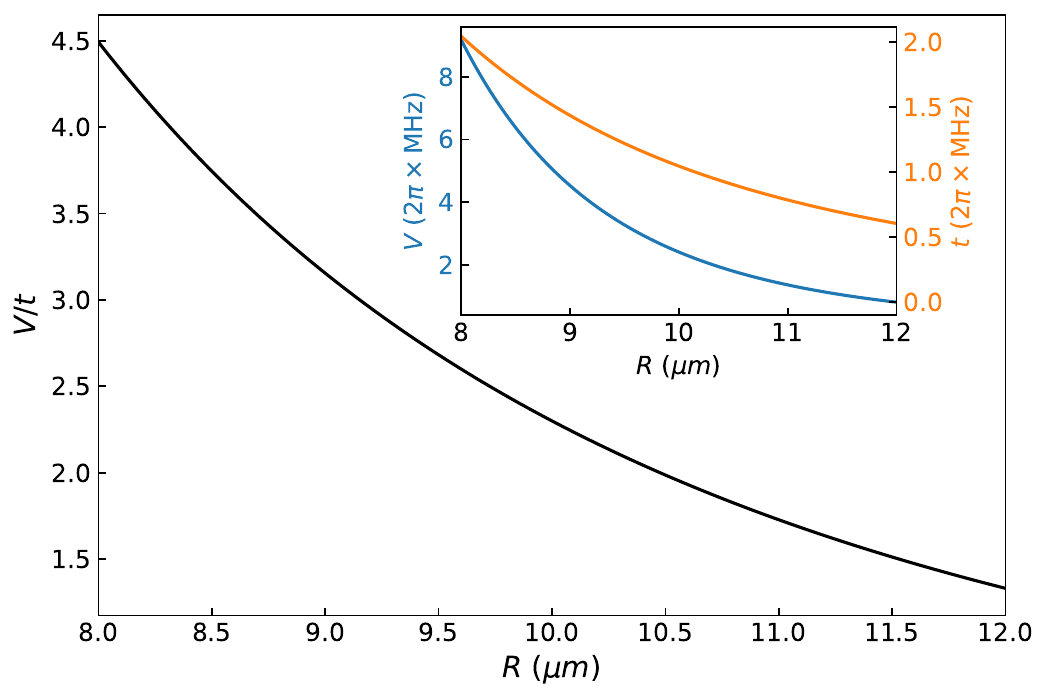}
   \caption{Calculated ratio $V/t$ for the proposed Rydberg states as a function of the interatomic distance $R$. The inset shows the corresponding interaction strengths $V$ and $t$ (in MHz).}
    \label{fig:S13}
\end{figure}
In two-dimensional Rydberg arrays, isotropic dipolar and van der Waals interactions are typically obtained by aligning the quantization axis perpendicular to the atomic plane ($\theta = 90^\circ$). In this standard configuration, the relevant microscopic parameters include the dipolar-assisted tunneling amplitude $t \propto C_{3}/R^{3}$ and the hole–hole interaction $V \propto C_{6}/R^{6}$, where $R$ is the lattice spacing and $C_{3}, C_{6}$ are the interaction coefficients dependent on the chosen Rydberg states and applied magnetic field.

While the geometric constraint inherently links $t$ and $V$, the experimentally relevant ratio $V/t$ remains highly controllable via the lattice spacing $R$ and the choice of Rydberg levels.  As a concrete example, we take the following Rydberg states of $^{87}$Rb atoms:
\[
\ket{78S_{1/2}, m_J=-\tfrac{1}{2}} \sim \ket{\downarrow},\qquad
\ket{79S_{1/2}, m_J=-\tfrac{1}{2}} \sim \ket{\uparrow},\qquad
\ket{78P_{3/2}, m_J=+\tfrac{1}{2}} \sim \ket{h}.
\]
We compute the tunneling strength $t=t_{\uparrow}$ (a slight spin dependence of the hole tunneling $t_{\uparrow}/t_{\downarrow} \approx 0.976$ for the chosen Rydberg states) and hole-hole interaction strength $V$ using the \textit{pairinteraction} package~\cite{Weber_2017} at a magnetic field of $B=75\,\mathrm{G}$ and $\theta = 90^\circ$. Fig.~\ref{fig:S13} illustrates their dependence on the interatomic spacing $R$. The ratio $V/t$ can be tuned across a broad window by varying $R$, highlighting the substantial flexibility available in current Rydberg tweezer platforms.  Furthermore, beyond static geometric tuning, highly independent control of the interaction terms is achievable via advanced  Hamiltonian engineering protocols. Techniques such as Floquet engineering of spin interactions~\cite{Geier_2021,PRXQuantum.3.020303} and microwave dressing schemes~\cite{PhysRevA.84.033619} allow for the independent adjustment of the XXZ interaction strengths, $J^z$ and $J^\perp$. Consequently, these methods effectively modify the interaction parameters, significantly expanding the accessible parameter space for  Rydberg quantum simulations. Finally, the rapid development of tweezer arrays with other atomic species (such as Cs, Na, K, Yb, and Sr) offers additional Rydberg series that potentially realize the target Hamiltonian, further broadening the experimental applicability of corresponding results.

\end{document}